\documentclass[preprint,amsmath,amssymb,aps,superscriptaddress]{revtex4-2}
\usepackage{amsmath}
\usepackage{textcase}
\usepackage{amssymb}
\usepackage{graphicx}
\usepackage{bm}
\usepackage{float}
\usepackage{multirow,microtype,color,relsize,ulem}
\usepackage{subfigure}
\usepackage[breaklinks=true]{hyperref}
\usepackage[utf8]{inputenc}
\usepackage[english]{babel}
\hypersetup{colorlinks=true,linkcolor=blue,citecolor=blue,urlcolor=black}
\hypersetup{linktocpage}
\usepackage{xr}
\usepackage{float}

\begin{document}
		\title{Floquet Prethermal Phase Protected by $U(1)$ Symmetry on a Superconducting Quantum Processor}
		\author{Chong Ying}
		\thanks{These authors contributed equally.}
		\affiliation{Hefei National Laboratory for Physical Sciences at the Microscale and Department of Modern Physics, University of Science and Technology of China, Hefei 230026, China}
		
		\affiliation{Shanghai Branch, CAS Center for Excellence in Quantum Information and Quantum Physics, University of Science and Technology of China, Shanghai 201315, China}
		
		\affiliation{Shanghai Research Center for Quantum Sciences, Shanghai 201315, China}
		
		\author{Qihao Guo}
		\thanks{These authors contributed equally.}
			
		\affiliation{Shenzhen Institute for Quantum Science and Engineering, Southern University of Science and Technology, Shenzhen 518055, China}
		
		\affiliation{Center for Quantum Technology Research and Key Laboratory of Advanced Optoelectronic Quantum Architecture and Measurements (MOE), School of Physics, Beijing Institute of Technology, Beijing 100081, China}
		
		\affiliation{Guangdong Provincial Key Laboratory of Quantum Science and Engineering Southern University of Science and Technology, Shenzhen, Guangdong 518055, China}
		
		\author{Shaowei Li}
		\thanks{These authors contributed equally.}
		\author{Ming Gong}\email{minggong@ustc.edu.cn}		
		\affiliation{Hefei National Laboratory for Physical Sciences at the Microscale and Department of Modern Physics, University of Science and Technology of China, Hefei 230026, China}
		
		\affiliation{Shanghai Branch, CAS Center for Excellence in Quantum Information and Quantum Physics, University of Science and Technology of China, Shanghai 201315, China}
		
		\affiliation{Shanghai Research Center for Quantum Sciences, Shanghai 201315, China}
		
		\author{Xiu-Hao Deng}
		
		\affiliation{Shenzhen Institute for Quantum Science and Engineering, Southern University of Science and Technology, Shenzhen 518055, China}
				
		\affiliation{Guangdong Provincial Key Laboratory of Quantum Science and Engineering Southern University of Science and Technology, Shenzhen, Guangdong 518055, China}

		\author{Fusheng Chen}
		\author{Chen Zha}
		\author{Yangsen Ye}
		\author{Can Wang}
		\author{Qingling Zhu}
		\author{Shiyu Wang}
		\author{Youwei Zhao}
		\author{Haoran Qian}
		\author{Shaojun Guo}
		\author{Yulin Wu}
		\author{Hao Rong}
		\author{Hui Deng}
		\author{Futian Liang}
		\author{Jin Lin}
		\author{Yu Xu}
		\author{Cheng-Zhi Peng}
		\author{Chao-Yang Lu}
		
		\affiliation{Hefei National Laboratory for Physical Sciences at the Microscale and Department of Modern Physics, University of Science and Technology of China, Hefei 230026, China}
		
		\affiliation{Shanghai Branch, CAS Center for Excellence in Quantum Information and Quantum Physics, University of Science and Technology of China, Shanghai 201315, China}
		
		\affiliation{Shanghai Research Center for Quantum Sciences, Shanghai 201315, China}
		
		\author{Zhang-Qi Yin}\email{zqyin@bit.edu.cn}
		
		\affiliation{Center for Quantum Technology Research and Key Laboratory of Advanced Optoelectronic Quantum Architecture and Measurements (MOE), School of Physics, Beijing Institute of Technology, Beijing 100081, China}
		
		\affiliation{Beijing Academy of Quantum Information Sciences, Beijing 100193, China}
		
		\author{Xiaobo Zhu}
		\author{Jian-Wei Pan}
		
		\affiliation{Hefei National Laboratory for Physical Sciences at the Microscale and Department of Modern Physics, University of Science and Technology of China, Hefei 230026, China}
		
		\affiliation{Shanghai Branch, CAS Center for Excellence in Quantum Information and Quantum Physics, University of Science and Technology of China, Shanghai 201315, China}
		
		\affiliation{Shanghai Research Center for Quantum Sciences, Shanghai 201315, China}
		
		\date{\today}
		
		\begin{abstract}
		Periodically driven systems, or Floquet systems, exhibit many novel dynamics and interesting out-of-equilibrium phases of matter. Those phases arising with the quantum systems' symmetries, such as global $U(1)$ symmetry, can even show dynamical stability with symmetry-protection. Here we experimentally demonstrate a $U(1)$ symmetry-protected prethermal phase, via performing a digital-analog quantum simulation on a superconducting quantum processor. The dynamical stability of this phase is revealed by its robustness against external perturbations. We also find that the spin glass order parameter in this phase is stabilized by the interaction between the spins. Our work reveals a promising prospect in discovering emergent quantum dynamical phases with digital-analog quantum simulators.
		\end{abstract}
		\maketitle
	\noindent
    \textbf{Introduction}\label{Introduction}: Searching for novel phases of matter is an eternal task in the field of condensed matter. 
    In traditional condensed matter theory, all the phases of equilibrium matter were described by Landau’s symmetry-breaking theory~\cite{landau1936theory} for a long time until the discovery of topological order ~\cite{wen2019choreographed,zeng2019quantum} broadened the range of states of matter. Recently, an evolution has happened in the field of far-from-equilibrium condensed matter~\cite{PhysRevLett.121.170501,smith2019simulating,bluvstein2021controlling}. The progress of driven quantum time-periodic systems, namely Floquet systems, has stimulated further interest in the search for far-from-equilibrium phases. A conventional view is that, in such a system, the information encoded on the initial state will be rapidly erased due to the inevitable thermalization induced by the continuous driving~\cite{PhysRevX.10.021046,PhysRevX.10.011043,ueda2020quantum,PhysRevX.10.021046}. Two important mechanisms have been considered to prevent the information loss and lead to long-lived phases under the Floquet drive: The first is many-body localization (MBL), in which the eigenstate thermalization hypothesis (ETH) falis~\cite{ueda2020quantum}; The other one is prethermalization, where the thermalization rate is exponentially small~\cite{ueda2020quantum,Peng2021,kyprianidis2021observation}. Phenomena of prethermalization have been studied in both on static systems~\cite{Gring1318,PhysRevA.89.033616,PhysRevLett.111.197203,PhysRevLett.115.180601} and Floquet systems~\cite{PhysRevX.7.011026,PhysRevB.95.014112,mori2016rigorous,kuwahara2016floquet,abanin2017rigorous}, whose properties can be captured by their effective static Hamiltonian. Most of the observed non-equilibrium long-lived behaviors~\cite{Zhang2017,Choi2017,kyprianidis2021observation,PhysRevLett.120.180603,PhysRevResearch.1.033202} are considered to be in the prethermal regime, which inspires a lot of interests in the search of the novel phases~\cite{PhysRevX.7.011026,PhysRevX.10.021046,ippoliti2020manybody,PhysRevA.99.033618,vogl2019flow,Peng2021}. Prethermal phases are generally featured by a quasi-stationary state with long-lived equilibrium-like properties~\cite{PhysRevX.9.021027,Abanin_2015,Abanin_2017,PhysRevX.10.021046,PhysRevX.10.011043,kyprianidis2021observation,Gring1318}. Typically, these phases can exist in the interacting quantum systems with various symmetries, such as a $U(1)$ symmetry~\cite{ueda2020quantum,PhysRevX.10.021046,PhysRevX.9.021027} and spatial symmetries~\cite{kaminishi2015entanglement,ueda2020quantum}.
    
    \begin{figure*}[htbp]
    	\centering
    	\includegraphics[width=0.7\textwidth]{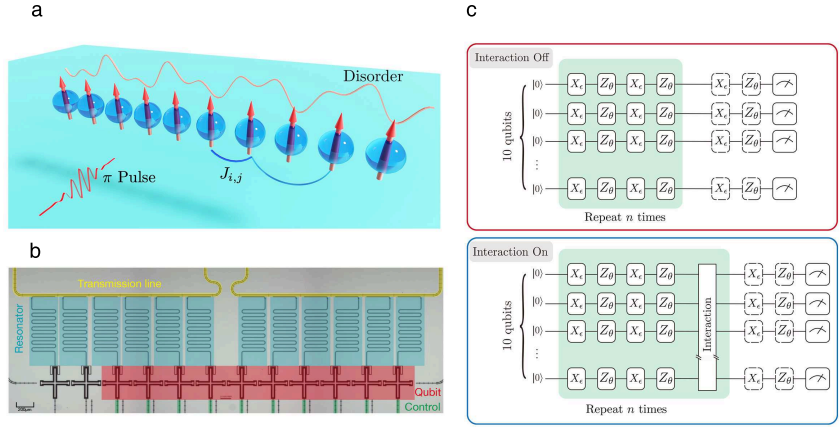}
    	\caption{\label{Fig.1}\textbf{Schematics of the experiment}. \textbf{a}, The illustration of the interacting Floquet system via digital-analog quantum simulation. The spin chain is initialized on the $\sigma_z$ basis fully polarized state. Three kinds of Hamiltonians are applied sequentially in time: a global imperfect flip Hamiltonian $ H_{\text{Flip}} $ ($\pi$~pulse), strong disorder $ H_{\text{Disorder}} $ and XX interaction $ H_{\text{Int}} $. \textbf{b}, The twelve transmon qubits, illustrated as black crosses, are arranged in a 1D chain. In our experiment, we choose $ 10 $ qubits in the right red region. The direct coupling between them is realized via capacitive coupling. Each qubit has individual Z and XY (green) control lines, and is coupled to corresponding readout resonators (blue) for dispersive readout. Six resonators in a group are coupled to a shared transmission line (yellow). \textbf{c}, The Floquet protocol of interacting Floquet matter with digital-analog simulation. Two cases, denoted by ``Interaction Off" and ``Interaction On", are shown in the red box and the blue box, respectively. Rotation-$ X $ gates and virtual $ Z $ gates in the circuits are used to simulate the the flip Hamiltonian $ H_{\text{Flip}} $ and the disorder Hamiltonian $ H_{\text{Disorder}} $. In a complete Floquet period, we apply two cycles of the flip Hamiltonian $ H_{\text{Flip}} $ and the disorder term $ H_{\text{Disorder}} $, and add an interaction Hamiltonian $ H_{\text{Int}} $ in the ``Interaction On" case. We measure the system after repeating complete Floquet periods $n$ times, where $ n \in [1,50] $. By replacing the last complete Floquet period of these sequences with a reduced cycle consisting of a flip term and a disorder term (dashed gates), we can measure the intermediate state of spins under one complete period, which results in ``Half Floquet Periods".}
    \end{figure*}
    
      Recent theoretical progresses have proven the existence of Floquet prethermal phases with the emergent symmetry-protection~\cite{PhysRevX.10.011043,PhysRevX.10.021046,Peng2021,PhysRevX.10.021046}. These phases may even exhibit extraordinary robustness against the transverse perturbations for a relatively long term. Experimental works show the existence of a quasi-conserved observable~\cite{Peng2021,PhysRevB.103.054305}, corresponding to the emergent symmetry ~\cite{Luitz_2020,doi:10.1146/annurev-conmatphys-031218-013721,khemani2019brief}. Intuitively, the quasi-conserved observable can be observed in a more sophisticated Floquet system whose symmetry is weakly broken. Moreover, beyond the static state approximation, whether a non-equilibrium phase possesses a non-trivial order parameter is still an open question. In addition to these fundamental problems, it is also very challenging to study these prethermal phases experimentally. Firstly, pure analog quantum simulators have very limited controllability and hence are too specific to adjust the simulated models. Besides, the scalability of quantum systems also sets a size limit for the quantum simulation. To tackle these issues, more universally-controllable quantum systems are desired. In the Noisy Intermediate Scale Quantum (NISQ) era of quantum computing, the digital-analog quantum simulation (DAQS)~\cite{PhysRevA.101.022305,PhysRevA.101.052337} is believed to be an efficient way to achieve this goal.
    
    In this Letter, we implement a $U(1)$ symmetry-protected prethermal phase on a superconducting quantum processor. The existence of dynamical quasi-conserved observable and the emergent $ U(1) $ symmetry protect the system. An increasing perturbative transverse field is applied on the Floquet matter to gradually break the emergent $U(1)$ symmetry. In sharp contrast to the non-interacting case, the dynamical quasi-conserved observable shows a collective quasi-steady behavior for a long term. We measure the dynamics of quasi-conserved observable and characterize the spectrum feature of Floquet matter by its Fourier spectrum. Furthermore, inspired by the similar feature between the interacting Floquet spin matter and the spin glass, we introduce the spin glass order parameter to characterize the intrinsic behavior in the system~\cite{Edwards_1975,PhysRevLett.113.107204,ippoliti2020manybody,RevModPhys.58.801}. With the correlated observables, we reconstruct the dynamics of spin glass order parameter and clarify the necessity of interaction to form a stable phase in such a disordered system. All of our results are based on the DAQS.
    
    \noindent
    \textbf{Model and Experimental Setting}\label{Model and Experimental Setting}: We perform our experiment on an array of $ L=10 $ superconducting transmon qubits with XX interaction~\cite{gong2020experimental}, shown in Fig.~\ref{Fig.1}a and Fig.~\ref{Fig.1}b. The Floquet Hamiltonian $ \hat{H} $ within a period $ T $, consisting of imperfect global flips, disorders on sites and the XX interaction, is implemented via digital-analog quantum simulation (See Fig.\ref{Fig.1}c) ($ \hbar =1 $):
    \begin{equation}\label{Eq1}
    	\hat{H} = \begin{cases}
    		H_{\text{Flip}}=g(1+\epsilon) \sum^L_{i=1} \sigma^x_i/2\,, & \text{time} \ t_1 \\
    		H_{\text{Disorder}}= \sum^{L}_{i=1} h_i \sigma^z_i/2\,, & \text{time} \ t_2 \\
    		H_{\text{Int}}= \sum^{L-l}_{i=1} J^{(l)}_i(\sigma^x_i\sigma^x_{i+l}+\sigma^y_i\sigma^y_{i+l})/2\,, & \text{time}\  t_3   
     \end{cases}\,.
    \end{equation}
    where $l=1,2$. Here, $ \sigma^\gamma_i $ is the Pauli operators acting on the $ i $th site. $ g $ is the Rabi frequency with small perturbation $ \epsilon $. $ h_i $ is a site-dependent disordered potential and $ J^{(l)}_i $ is the coupling strength between site $ i $ and site $ (i+l) $. The capacitive coupling leads to the nearest-neighbor coupling strength $ J^{(1)}_i $, with the average value of $ J^{(1)}_i  \sim 2\pi\times 10.84 $~MHz, and the next-nearest-neighbor coupling strength $ J^{(2)}_i $, with the average value of $ J^{(2)}_i \sim 2\pi \times 0.28 $~MHz $\ll J^{(1)}_i $.
    
     The Floquet protocol of system is $ H_{\text{Flip}} \to H_{\text{Disorder}} \to H_{\text{Flip}} \to H_{\text{Disorder}} \to H_{\text{Int}} $ (as shown in Fig.\ref{Fig.1}\textbf{c}), which is similar to discrete time crystal (DTC) experiments~\cite{Zhang2017,Choi2017,kyprianidis2021observation,PhysRevLett.120.180603} over two-period under the first order Mangus expansion~\cite{Luitz_2020,doi:10.1002/cmr.a.21414,supp}. The unitary time evolution $ U $ in the experiment under a single Floquet period $ T = 2t_1 +2t_2+ t_3$ is
    \begin{equation}\label{Eq2}
    	U(T) = e^{-iH_{\text{Int}} t_3} (e^{-i H_{\text{Disorder}} t_2} X_{\epsilon})^2 \,.
    \end{equation}
    where $ X_{\epsilon}$ is the imperfect flip operator generated by the flip Hamiltonian $ H_{\text{Flip}} $. With the XX interaction $H_{\text{Int}}$, which has $U(1)$ symmetry, the effective Hamiltonian of the Floquet operator $U(T)$ is considered to be a Floquet prethermal Hamiltonian with a weakly symmetry-breaking term $\epsilon$~\cite{PhysRevX.9.021027,PhysRevX.10.021046,khemani2019brief}. In addition, a Floquet system with the absence of $H_{\text{Int}}$ is implemented as `` Interaction Off' case to study the the contribution of spin interaction in producing the prethermal phase of the Floquet matter. In comparison with the ``Interaction On" case, there is no Floquet prethermalization in the `` Interaction Off' case. 

    In order to generate the Floquet Hamiltonian, we implement the flip Hamiltonian $ H_{\text{Flip}} $ and the disorder Hamiltonian $ H_{\text{Disorder}} $ with digital gates, and the interaction Hamiltonian $ H_{\text{Int}} $ with analog quantum simulation. The $ X_{\epsilon}$ operator flips all the spins around the $ x $-axis of the Bloch sphere with a controllable perturbation $ \epsilon $. We optimize control pulses with Derivative Removal via Adiabatic Gate (DRAG) to suppress the state leakage out of the computational basis during the $X_{\epsilon}$ flip operation\cite{PhysRevA.82.040305}, and the average fidelity of simultaneous single qubit gate is estimated via randomized benchmarking (RB) yielding around $99.4\%$. The disorder term $  e^{-iH_{\text{Disorder}}t_2} $ is synthesized by the residual frequencies derived from the rotating frame and additional virtual $Z$ gates~\cite{supp,PhysRevA.96.022330}. To design a random and tunable disorder, we separately apply virtual $Z$ gates on each qubit with random rotation angles $ \theta_i$ which is drawn from a uniform distribution $ \theta_i \sim [-\pi,\pi] $. The interaction Hamiltonian $ H_{\text{Int}} $ is achieved by detuning all qubits on resonance, where the imperfect frequency drifts are corrected via multi-qubit excitation propagation method.
    
    Finally, we measure the system on the $\sigma_z$ basis. To probe the dynamical behavior of spins under one Floquet period, we introduce a ``Half Floquet Period" measurement: replacing the last complete Floquet period of the repeated complete Floquet periods with a reduced cycle consisting of the flip operation $ X_\epsilon$ and the disorder operator $  e^{-iH_{\text{Disorder}}t_2} $ (shown as the dashed quantum gates in Fig.~\ref{Fig.1}\textbf{c}). The evolution time $t$ after the complete Floquet cycles $n$ meets $t = nT$. For the ``Half Floquet Period" $\tilde{n}$, it is given by $t = (t_1 +t_2)* (\tilde{n} - 2*[\tilde{n}/2]) + T*[\tilde{n}/2]$, where $[x]$ is the floor function which returns the integral part of nonnegative number $x$. We reconstruct the population of each qubit and the correlator $\langle \sigma_i \sigma_j\rangle$ between two qubits for all the ``Half" and complete Floquet cycles.
    
    \begin{figure*}[htbp]
     \centering
    \includegraphics[width=0.7\textwidth]{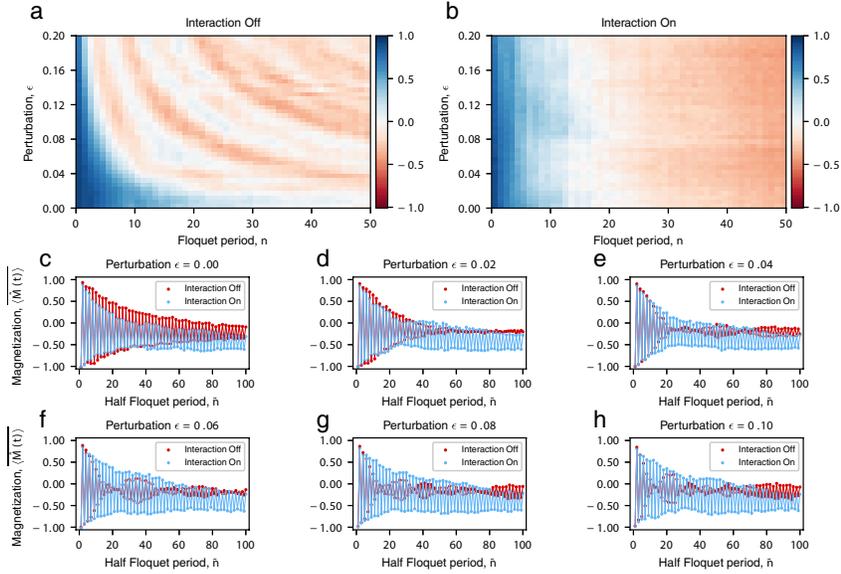}
    \caption{\label{Fig.2}\textbf{Dynamics of the Floquet System with Various Perturbation $ \epsilon $}. \textbf{a}, in the ``Interaction Off" case, the phase structure of $ \overline{\langle\hat{M}(t) \rangle} $ varies with different perturbation strength $ \epsilon $. When the external perturbation increases, $ \overline{\langle\hat{M}(t) \rangle} $ of the system will flip with a shorter period. \textbf{b}, in the ``Interaction On" case, the phase structure of $ \overline{\langle\hat{M}(t) \rangle} $ is not sensitive to the perturbation strength $ \epsilon $ as large as $0.2$. \textbf{c}-\textbf{h}, the dynamics of  $ \overline{\langle\hat{M}(t) \rangle} $. We initialize the system in $ |0\rangle^{\otimes 10} $ state and measure the system in ``Half Floquet Periods", of which even periods corresponds to the complete Floquet periods. Each measurement has been repeated for $2000$ times. Average error bars in the figure are estimated about $ 0.01 $.}
    \end{figure*}
     \noindent
    \textbf{Results}: To study the Floquet prethermal phase generated by the Floquet operator $U(T)$, we now consider two important features of the Floquet system, the spatially-averaged spin magnetization $ \overline{\langle\hat{M}(t)\rangle}  $ with its Fourier spectrum $  P(\omega)$~\cite{Zhang2017,sahay2020emergent,ippoliti2020manybody}, and the spin glass order parameter $  \chi_{\text{SG}} $~\cite{Edwards_1975,Khemani_2016,ippoliti2020manybody,RevModPhys.58.801,PhysRevB.99.144201}, which are associated with temporal correlation function and spatial correlation function respectively.

     We first consider the dynamics of spatially-averaged spin magnetization $ \overline{\langle\hat{M}(t)\rangle}=\frac{1}{L}\sum^L_i \langle \hat{\sigma}^z_i (t) \rangle $, which is the quasi-conserved observable of the $U(1)$ symmetry. To clarify the role of $U(1)$ symmetry, we add an external transverse field $\epsilon\sum^L_{i=1}\sigma^x_i/2$ to break the symmetry. Fig.~\ref{Fig.2}\textbf{a}-\textbf{b} display the phase structure of $\overline{\langle\hat{M}(t)\rangle}$ varying with the external perturbation $ \epsilon $, up to $ 50 $ complete Floquet periods. In ``Interaction Off" case and ``Interaction On" case, it appears different responses to the transverse field. Without the interaction Hamiltonian $ H_{\text{Int}} $, the dynamics of the magnetization is sensitive to the external perturbation $ \epsilon $, while the XX interaction eliminates the effect of perturbation $\epsilon$. In addition, we observed obvious streaks in Fig.~\ref{Fig.2}\textbf{a}, which shows the dramatic change of $\overline{\langle\hat{M}(t)\rangle}$ in the ``Interaction Off" case. By contrast, the dynamics of $\overline{\langle\hat{M}(t)\rangle}$ of the ``Interaction On" is robust to the external perturbation in Fig.~\ref{Fig.2}\textbf{b}. With the flip Hamiltonian $H_{\text{Flip}}$, the system will flip two times in a single Floquet period as a non-equilibrium matter, which can be measured in each ``Half Floquet Periods". We investigate the dynamics of both ``Interaction Off" and ``Interaction On" cases, shown as Fig.~\ref{Fig.2}\textbf{c}-\textbf{h}. The existence of interaction makes the dynamics distinct from ``Interaction Off" case, which does not only prolong the oscillation time but also prevents the formation of beats on the envelope. When $ \epsilon= 0 $, both ``Interaction Off" and ``Interaction On" cases show long-lived oscillation without beats. However, when we add the symmetry-broken term $ g \epsilon\sum_i \sigma^x_i$ into the system, non-interacting system will oscillate with the beats and rapidly  decay due to the external perturbation. In contrast, the interacting system still keep the long-lived behavior, which implies a non-trivial symmetry-protection mechanism in the system. Theoretically, $ \overline{\langle \hat{M}(t) \rangle} $ in the interacting system, as a Floquet quasi-conserved observable (FQO), show a long-lived behavior derived from the approximate $U(1)$ symmetry~\cite{PhysRevX.10.021046}. Previous work reported that the FQO was observed in a $0$ spin-glass-like Floquet prethermal phase~\cite{khemani2019brief} with a uniform longitude field and dipolar interaction~\cite{Peng2021}, whose magnetization quickly converges to a non-zero value as a prethermal plateau. In sharp contrast to their work, our system does not only emerge a long-lived prethermal phenomenon, but also shows that the approximate $U(1)$ symmetry can protect a $\pi$ spin-glass-like prethermal phase whose magnetization always keeps oscillating as a dynamical FQO. The background decay of $ \overline{\langle \hat{M}(t) \rangle} $ displayed in Fig.~\ref{Fig.2}\textbf{c} results from the effects of the qubit decoherence, the leakage error and the flip gate error~\cite{supp}.
    \begin{figure}[htbp]
	\centering
	\includegraphics[width=0.7\textwidth]{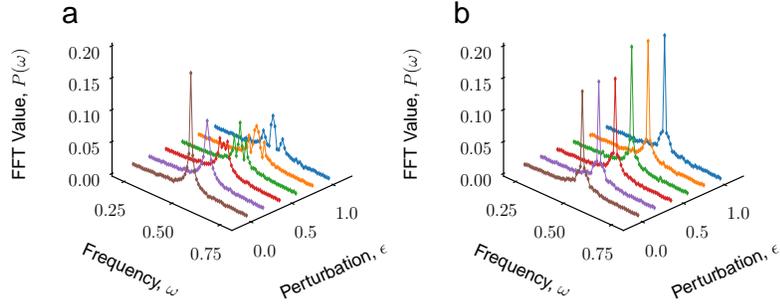}
	\caption{\label{Fig.3} \textbf{Spectra of the spatially-averaged spin magnetization, $ P(\omega)$}. We investigate the Fourier spectrum of the Floquet matter with the external perturbation in ``Half Floquet Periods" measurements. \textbf{a}, in the ``Interaction Off" case, when the strength of perturbation $\epsilon$ is increasing, the height of Fourier spectrum gradually decays and then the peak in the spectrum center splits into three peaks.
	\textbf{b}, in the ``Interaction On" case, only a peak emerges in the Fourier spectrum and the height of it is stable. The great difference between \textbf{a} and \textbf{b} characterizes the different spectrum features of the non-interacting matter and interacting matter.}
\end{figure}

    Then we investigate the spectrum feature of the Floquet matter with the external perturbation $ \epsilon $, which weakly breaks the $U(1)$ symmetry. The Fourier spectrum of $ \overline{\langle \hat{M}(t) \rangle} $ characterizes the frequency response of the matter under the external field. Fig.~\ref{Fig.3} shows the great difference between the spectra of``Interaction Off" case and the ``Interaction On" case, which are extracted from $ \overline{\langle \hat{M}(t) \rangle} $ measured in each ``Half Floquet Periods". For the non-interacting system whose spins are continuously flipped by the imperfect $ \pi $ pulse, the Fourier spectrum $  P(\omega)$ is not robust with the presence of disorders, formally split into many peaks~\cite{Zhang2017,supp}. However, with the existence of the interaction, only a single peak appears on the Fourier spectrum, which means the response frequency of the Floquet prethermal phase is robust to the external perturbation. The behavior is similar to the DTC system but does not have a rigorous sub-harmonic response ~\cite{PhysRevB.97.184301,PhysRevLett.120.180603,PhysRevA.99.033618}. 
  \begin{figure}[htbp]
	\centering
	\includegraphics[width=0.7\textwidth]{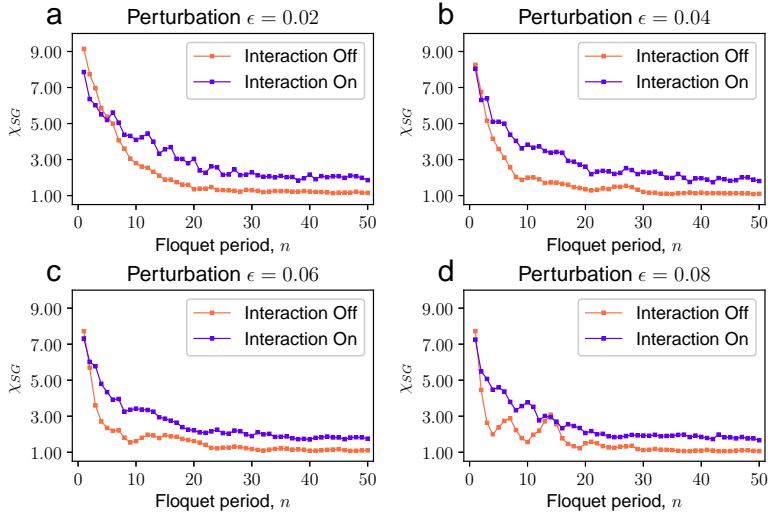}
	\caption{\label{Fig.4} \textbf{Dynamics of the spin glass order parameter $  \chi_{\text{SG}} $}. \textbf{a}-\textbf{d}, The spin glass order parameter $  \chi_{\text{SG}} $ as function of external perturbation $ \epsilon $, averaged over each period and repeated $ 2000 $ times. The ``Interaction Off" case (the orange curve) decays faster than the ``Interaction On" case (the purple curve), which implies the necessity of interaction in the system to stabilize the system.}
\end{figure}

    The collective behavior of the Floquet matter can be investigated with $ \overline{\langle\hat{M}(t)\rangle} $ and $P(\omega)$. However, they cannot be used to confirm the intrinsic ordered structure of the system. Previous works show that the localization-protected quantum order parameter ~\cite{PhysRevB.88.014206,PhysRevB.99.144201,PhysRevLett.113.107204} can be an appropriate order parameter in a spin-glass-like system. For instance, the spin glass order parameter $  \chi_{\text{SG}} $~\cite{PhysRevLett.113.107204,PhysRevLett.125.240401}, giving
    \begin{equation}\label{key}
    	\begin{aligned}
        	 \chi_{\text{SG}}(t) &= \frac{1}{L} \sum_{i,j} \langle \psi_0| \hat{\sigma}^z_i (t) \hat{\sigma}^z_j (t) | \psi_0\rangle^2\\
    	& = 1 + \frac{2}{L} \sum^L_{i<j}  \langle \psi_0| \hat{\sigma}^z_i (t) \hat{\sigma}^z_j (t) | \psi_0\rangle^2\,,
    	\end{aligned}
    \end{equation}
	is predicted to be extensive in a quantum phase with glassy order, like DTCs~\cite{ippoliti2020manybody}. Here we probe the two-spin equal-time correlation function $ \langle \sigma^z_i \sigma^z_j \rangle$ ($ i \neq j $) in the Floquet matter and reconstruct $ \chi_{\text{SG}} $. Fig.~\ref{Fig.4} illustrates behaviors of $ \chi_{\text{SG}} $ with different external perturbation $ \epsilon $. When the interaction is off, the spin correlation will rapidly vanish due to the disturbance of random disorders. When the interaction is on, the spins are locked into the same direction, and the decay of $  \chi_{\text{SG}} $ appears slower. The slow vanishment of $\chi_{\text{SG}} $ reveals the relatively long lifetime of the intrinsic ordered structure in the Floquet prethermal phase. For further verification, we numerically calculate the order under an XX-interaction and an Ising-interaction system~\cite{supp}. 

     \noindent
    \textbf{Discussion}: In conclusion, we have proposed and experimentally realized a prethermal phase of the interacting Floquet matter, which has an emerged $ U(1) $ symmetry with a dynamical quasi-conserved observable. We showed that the approximate $ U(1) $ conservation protects the prethermal phase for a long time, and further revealed the difference of spectrum features of the Floquet systems under the external perturbation. Then we found that the existence of the interaction entirely changed the behavior of the spin glass order parameter, making its decay slower. In fact, looking for appropriate order parameters for the prethermal phases with various emergent symmetries is still an open problem. These results suggest that the approximate symmetries in Floquet systems provide dynamical stability for the prethermal phases. On the other hand, the phases of matter beyond equilibrium states can be greatly enriched by applying various novel Floquet protocols~\cite{doi:10.1146/annurev-conmatphys-031218-013721,von2016phase}. In particular, recent works reported that such prethermal phases can exist even in the thermodynamics limit $L\to \infty$ and a prethermal DTC was observed~\cite{ye2021classical, pizzi2021classical,kyprianidis2021observation}. Our work motivates further investigations to the Floquet many-body system and indicates a route towards effective quantum simulation with DAQS in the field of quantum many-body physics.

\begin{acknowledgments}
	\noindent The authors thank the USTC Center for Micro- and Nanoscale Research and Fabrication. The authors also thank QuantumCTek Co., Ltd. for supporting the fabrication and maintenance of the room temperature electronics. This research was supported by the National Key R$\&$D Program of China (No. 2017YFA0304300), the Key-Area Research and Development Program of Guangdong Province (Grant No. 2020B0303030001 and 2018B030326001), the Chinese Academy of Sciences, and Shanghai Municipal Science and Technology Major Project (Grant No. 2019SHZDZX01), the National Natural Science Foundation of China (No. 61771278, U1801661, and 11905217), Beijing Institute of Technology Research Fund Program for Young Scholars, the Guangdong Provincial Key Laboratory (Grant No.2019B121203002), the Natural Science Foundation of Guangdong Province (2017B030308003), Anhui Initiative in Quantum Information Technologies, the Guangdong Innovative and Entrepreneurial Research Team Program (2016ZT06D348), the Science, Technology and Innovation Commission of Shenzhen Municipality (JCYJ20170412152620376, KYTDPT20181011104202253), and the NSF of Beijing (Grants No. Z190012).
\end{acknowledgments}

	\bibliography{apssamp}
\end{document}


\setcounter{equation}{0}
	\setcounter{figure}{0}
	\setcounter{table}{0}
	\setcounter{page}{1}
	\makeatletter
	\renewcommand{\theequation}{S\arabic{equation}}
	\renewcommand{\thefigure}{S\arabic{figure}}
	\renewcommand{\bibnumfmt}[1]{[S#1]}
	\renewcommand{\citenumfont}[1]{S#1}
	
	\title{Supplemental Materials: Floquet Prethermal Phase Protected by $U(1)$ Symmetry on a Superconducting Quantum Processor}
	\author{Chong Ying}
		\thanks{These authors contributed equally.}
		\affiliation{Hefei National Laboratory for Physical Sciences at the Microscale and Department of Modern Physics, University of Science and Technology of China, Hefei 230026, China}
		
		\affiliation{Shanghai Branch, CAS Center for Excellence in Quantum Information and Quantum Physics, University of Science and Technology of China, Shanghai 201315, China}
		
		\affiliation{Shanghai Research Center for Quantum Sciences, Shanghai 201315, China}
		
		\author{Qihao Guo}
		\thanks{These authors contributed equally.}
			
		\affiliation{Shenzhen Institute for Quantum Science and Engineering, Southern University of Science and Technology, Shenzhen 518055, China}
		
		\affiliation{Center for Quantum Technology Research and Key Laboratory of Advanced Optoelectronic Quantum Architecture and Measurements (MOE), School of Physics, Beijing Institute of Technology, Beijing 100081, China}
		
		\affiliation{Guangdong Provincial Key Laboratory of Quantum Science and Engineering Southern University of Science and Technology, Shenzhen, Guangdong 518055, China}
		
		\author{Shaowei Li}
		\thanks{These authors contributed equally.}
		\author{Ming Gong}\email{minggong@ustc.edu.cn}
		
		\affiliation{Hefei National Laboratory for Physical Sciences at the Microscale and Department of Modern Physics, University of Science and Technology of China, Hefei 230026, China}
		
		\affiliation{Shanghai Branch, CAS Center for Excellence in Quantum Information and Quantum Physics, University of Science and Technology of China, Shanghai 201315, China}
		
		\affiliation{Shanghai Research Center for Quantum Sciences, Shanghai 201315, China}
		
		\author{Xiu-Hao Deng}
		
		\affiliation{Shenzhen Institute for Quantum Science and Engineering, Southern University of Science and Technology, Shenzhen 518055, China}
				
		\affiliation{Guangdong Provincial Key Laboratory of Quantum Science and Engineering Southern University of Science and Technology, Shenzhen, Guangdong 518055, China}

		\author{Fusheng Chen}
		\author{Chen Zha}
		\author{Yangsen Ye}
		\author{Can Wang}
		\author{Qingling Zhu}
		\author{Shiyu Wang}
		\author{Youwei Zhao}
		\author{Haoran Qian}
		\author{Shaojun Guo}
		\author{Yulin Wu}
		\author{Hao Rong}
		\author{Hui Deng}
		\author{Futian Liang}
		\author{Jin Lin}
		\author{Yu Xu}
		\author{Cheng-Zhi Peng}
		\author{Chao-Yang Lu}
		
		\affiliation{Hefei National Laboratory for Physical Sciences at the Microscale and Department of Modern Physics, University of Science and Technology of China, Hefei 230026, China}
		
		\affiliation{Shanghai Branch, CAS Center for Excellence in Quantum Information and Quantum Physics, University of Science and Technology of China, Shanghai 201315, China}
		
		\affiliation{Shanghai Research Center for Quantum Sciences, Shanghai 201315, China}
		
		\author{Zhang-Qi Yin}\email{zqyin@bit.edu.cn}
		
		\affiliation{Center for Quantum Technology Research and Key Laboratory of Advanced Optoelectronic Quantum Architecture and Measurements (MOE), School of Physics, Beijing Institute of Technology, Beijing 100081, China}
		
		\affiliation{Beijing Academy of Quantum Information Sciences, Beijing 100193, China}
		
		\author{Xiaobo Zhu}
		\author{Jian-Wei Pan}
		
		\affiliation{Hefei National Laboratory for Physical Sciences at the Microscale and Department of Modern Physics, University of Science and Technology of China, Hefei 230026, China}
		
		\affiliation{Shanghai Branch, CAS Center for Excellence in Quantum Information and Quantum Physics, University of Science and Technology of China, Shanghai 201315, China}
		
		\affiliation{Shanghai Research Center for Quantum Sciences, Shanghai 201315, China}
		
		\date{\today}
	\maketitle
	\section{\label{sec:SQP}1. Superconducting Quantum Processor}
	We perform the experiment on a superconducting quantum processor with 12 transmon qubits in a 1D array with fixed coupling.
	Each qubit is controlled by a single line which combines both microwave driving and flux bias, and coupled to a resonator for dispersive readout.
	Six resonators in a group are coupled to a shared transmission line and can be measured simultaneously by frequency multiplexing technology.
	After the output port of each transmission line is connected an impedance-transformed parametric amplifier (IMPA) for improvement of the signal noise ratio. In addition, the wiring setup of the dilution refrigerator and electronics is the same as reported in Ref.~\citenum{zhu2021observation}.
	
	Due to the frequency crowding issue, 10 qubits on the processor are used in our experiment, with the other 2 qubits detuned far away by large flux bias.
	The coupling strengths between nearest-neighbor qubits and next-nearest-neighbor qubits measured at the working frequency $ \omega^{\text{work}}/2\pi = 4.15 $~GHz yield averages of $J_1/2\pi = 10.84$ MHz and $J_2/2\pi = 0.28$ MHz, respectively.
	
	\begin{table}[h]
	\centering
	\begin{tabular}{@{}lcccccccccccc@{}}
	\toprule
	\toprule
			& Q01& Q02& Q03 & Q04 & Q05 & Q06 & Q07 & Q08 & Q09 & Q10 \\
			\midrule
			$ \omega_{\text{Max}}/2\pi $~(GHz)  & 4.682& 4.646& 4.587& 4.622& 4.398& 4.486& 4.481& 4.621& 4.615& 4.550\\
			$ \omega_{\text{Idle}}/2\pi $~(GHz) & 4.066& 4.523& 3.773& 4.363& 3.853& 4.464& 3.963& 4.561& 3.883& 4.408\\
			$ \omega_{r}/2\pi $~(GHz) & 6.505& 6.535& 6.568& 6.599& 6.631& 6.652& 6.683& 6.714& 6.740& 6.767\\
			$ T_1 $~($ \mu $s)    & 82.9 & 26.9 & 50.1 & 37.3 & 72.5 & 68.4 & 24.1 & 70.6 & 71.0 & 36.6 \\
			$ T^*_2 $~($ \mu $s)    &0.79 & 1.78 & 1.08 & 1.60 & 0.90 & 4.16 & 0.69 & 2.42 & 0.79 & 2.32 \\
			$ \eta/2\pi $~(MHz)   &-252 & -244 & -256 & -246 & -256 & -244 & -254 & -244 & -256 & -248 \\
			$ g/2\pi $~(MHz) & 52.3 & 52.8 & 54.8 & 55.2 & 58.8 & 54.3 & 57.8 & 61.7 & 55.9 & 61.0\\
			$ \kappa_{r}^{-1} $~(ns) & 323 & 337 & 436 & 446 & 383 & 277 & 434 & 356 & 355 & 401\\
			$ f_{00} $ (\%)  &93.6& 97.0& 92.4& 93.9& 90.2& 96.5& 93.7& 95.5& 92.5& 95.7\\
			$ f_{11} $ (\%)   & 86.0& 86.9& 83.4& 85.3& 78.3& 90.2& 83.4& 86.1& 82.1& 87.7\\
	\bottomrule
    \bottomrule
	\end{tabular}
    \caption{\label{TableS1} \textbf{Parameters of our device}. $ \omega_{\text{Max}}/2\pi $ is the maximum frequency of the qubit. $ \omega_{\text{Idle}}/2\pi$ is the idle frequency of the qubit. $ \omega_{r} $ is the the frequency of the readout resonator. $ T_1 $ and $ T^*_2 $ are the energy relaxation and dephasing time of the qubit, respectively, which are both measured at the idle frequency. $ \eta/2\pi $ is the anharmonicity of qubit measured at the idle frequency. $ g/2\pi $ is qubit-resonator coupling strength. $ \kappa_{r}^{-1} $ is the leakage rate of readout resonators for characterization of resonator-transmission line coupling strength. $ f_{00} $ ($ f_{11} $) is the probability of correctly identifying the qubit state when it is initially prepared in $ |0\rangle $ ($ |1\rangle  $) state with simultaneous measurement. }
	\end{table}

	\section{\label{EI}2. Experiment Implementation}
In our system, the frequencies of qubits can be tuned by flux bias, which is contributed by DC signals and $ Z $ pulse. We utilize DC signals to set qubits at appropriate idle frequencies, and $ Z $ pulse to detune qubits from idle frequencies to working frequencies. Then we synchronize timing for $ XY $ microwave driving and $ Z $ pulse of all qubits. We calibrate qubits at their idle frequencies and benckmark the fidelity of single-qubit $ X_{\pi/2} $ gate simultaneously using randomized benchmarking (RB). Before detuning qubits to working frequency, we calibrate $ Z $ pulse crosstalk between each qubit pair and $ Z $ pulse distortion of each qubit. However there are still considerable frequency drifts when detuning all qubits to working frequency because of the imperfect calibration of $ Z $ pulse crosstalk.
Therefore, we use multi-qubit excitation propagation method to correct the frequency drifts. After correction the maximum of absolute frequency drifts is below $ 1 $ MHz.

In the experiment we need adding perturbation on $X$ gate, so we implement the sequence of \{$Y_{-\pi/2},Z_\epsilon,Y_{\pi/2}$\} in place of $X_\epsilon$ gate for the convenience of control. The pulse length of $Y_{-\pi/2}$ and $Y_{\pi/2}$ gates is $ 20 $ ns. We use virtual-$ Z $ strategy so that $Z$ gate takes zero time.
Therefore the effective pulse length of $X_\epsilon$ gates is $ 40 $ ns.
The pulse length of interaction detuing is $ 10  $~ns, including the ramp duration of $ 8 $~ns.

We repeat the experiment circuit and measurement every $ 200  $~$ \mu $s, ensuring the qubits decay to ground states. The probabilities of measured state are corrected with the product of calibration matrices constructed by simultaneous $ f_{00} $ and $ f_{11} $ of each qubit.

\section{3. System Hamiltonian and evolution}
The experimental qubit Hamiltonian of our system in the lab frame is 
\begin{equation}
	H = \begin{cases}
		H_{0}^{\text{idle}}+{{H}_{d}}(t)\,, & \text{time} \ t_1 \\
		H_{0}^{\text{work}}+{{H}_\text{XX}}\,, & \text{time}\  t_3
     \end{cases}\,,
\end{equation}
where
\begin{equation}
     H_{0}^{\text{idle}}=\sum\limits_{i=1}^{L}{\frac{\omega _{i}^{\text{idle}}}{2}\sigma _{i}^{z}}\,,
\end{equation}
\begin{equation}
    {{H}_{d}}(t)=\sum\limits_{i=1}^{L}{{{\Omega }_{i}}\left[ \cos (\omega _{i}^{\text{idle}}t)\frac{\sigma _{i}^{x}}{2}+\sin (\omega _{i}^{\text{idle}}t)\frac{\sigma _{i}^{y}}{2} \right]}\,,
\end{equation}
\begin{equation}
     H_{0}^{\text{work}}=\sum\limits_{i=1}^{L}{\frac{{{\omega }^{\text{work}}}}{2}\sigma _{i}^{z}}\,,
\end{equation}
\begin{equation}
     {{H}_\text{XX}}=\sum\limits_{n=1}^{2}{\sum\limits_{i=1}^{L-n}{\frac{J_{i}^{(n)}}{2}\left( \sigma _{i}^{x}\sigma _{i+n}^{x}+\sigma _{i}^{y}\sigma _{i+n}^{y} \right)}}\,.
\end{equation}

We use two different rotation operators in correlated stages according to qubits frequencies to find the transformed Hamiltonian
\begin{equation}
     \widetilde{H} = \begin{cases}
        \widetilde{H}_{\text{Flip}}^{\text{idle}}={{e}^{iH_{0}^{\text{idle}}t}}{{H}_{d}}(t){{e}^{-iH_{0}^{\text{idle}}t}}=\sum\limits_{i=1}^{L}{\dfrac{{{\Omega }_{i}}}{2}\sigma _{i}^{x}}\,, & \text{time} \ t_1 \\
        \widetilde{H}_{\text{Int}}^{\text{work}}={{e}^{iH_{0}^{\text{work}}t}}{{H}_{\text{XX}}}{{e}^{-iH_{0}^{\text{work}}t}}={{H}_{\text{XX}}}  \,, & \text{time}\  t_3
    \end{cases}\,.
\end{equation}

Without loss of generality, assuming the same initial state in different frames, we can find the state evolution in the rotating frame at working frequencies in the flip stage is
\begin{equation}
    \begin{aligned}
    {{\left| \widetilde{\psi }({{t}_{1}}) \right\rangle }^{\text{work}}}
    &={{e}^{iH_{0}^{\text{work}}{{t}_{1}}}}\left| \psi ({{t}_{1}}) \right\rangle \\
    &={{e}^{i(H_{0}^{\text{work}}-H_{0}^{\text{idle}}){{t}_{1}}}}{{\left| \widetilde{\psi }({{t}_{1}}) \right\rangle }^{\text{idle}}} \\
    &={{e}^{i(H_{0}^{\text{work}}-H_{0}^{\text{idle}}){{t}_{1}}}}{{e}^{-i\widetilde{H}_{\text{Flip}}^{\text{idle}}{{t}_{1}}}}\left| \psi (0) \right\rangle\,.
    \end{aligned}
\end{equation}

Now we only consider the evolution in the rotating frame at working frequencies under a single Floquet period $ T=2t_1+t_3 $. Between the flip stage and the interaction stage, we apply virtual $Z$ gates\cite{PhysRevA.96.022330} with different angle $\theta_i$ on each qubit. Therefore the unitary time evolution is
\begin{equation}
    U(T)={{e}^{-i\widetilde{H}_{\text{Int}}^{\text{work}}{{t}_{3}}}}{{\left( e^{-i\sum^L_i \theta_i \sigma^z_i/2} {{e}^{i(H_{0}^{\text{work}}-H_{0}^{\text{idle}}){{t}_{1}}}}{{e}^{-i\widetilde{H}_{\text{Flip}}^{\text{idle}}{{t}_{1}}}} \right)}^{2}}={{e}^{-i{{H}_{\text{XX}}}{{t}_{3}}}}{{\left( {{Z}_{\theta_i }}{{X}_{\epsilon }} \right)}^{2}}\,,
\end{equation}
where ${Z}_{\theta_i }$ is the disorder evolution operator and the effective disorder strength satisfies $h_i t_2 = (\theta_i-\omega^{\text{work}}t_1+\omega_i^{\text{idle}}t_1)/2 \mod 2\pi$.

	\section{\label{sec:MEFS}4. Magnus Expansion of Floquet Systems}
	Generally, the dynamics of Floquet system is different from the corresponding static system due to the non-commuting term in the Hamiltonian. We can analyse the behavior of a Floquet evolution operator via Magnus expansion\cite{doi:10.1002/cmr.a.21414,d2013many,PhysRevResearch.1.033202}. Firstly we considering the time evolution in Eq.~2:
	\begin{equation}\label{key}
		\begin{aligned}
			    	U(T) &= e^{-iH_{\text{Int}} t_3} (e^{-iH_{\text{Disorder}} t_2} e^{-iH_{\text{Flip}} t_1})^2 \,.\\
			    	&= e^{-iH_{\text{Int}} t_3}  e^{-iH_{\text{Disorder}} t_2} e^{- i\pi/2 \sum^L_{i}  \sigma^x_i} e^{- i\epsilon \pi/2 \sum^L_{i}  \sigma^x_i} e^{-iH_{\text{Disorder}} t_2} e^{- i\pi/2 \sum^L_{i}  \sigma^x_i} e^{- i\epsilon \pi/2 \sum^L_{i}  \sigma^x_i}\\
			    	& = (-1)^{L}e^{-iH_{\text{Int}} t_3}  e^{-iH_{\text{Disorder}} t_2} e^{- i\epsilon \pi/2 \sum^L_{i}  \sigma^x_i} e^{+iH_{\text{Disorder}} t_2} e^{- i\epsilon \pi/2 \sum^L_{i}  \sigma^x_i}\,, \\
		\end{aligned}
	\end{equation}
	where we used $ \{\sigma^z_i,\sigma^x_i\} =0 $ to eliminate the flip operator. Then we take the average Hamiltonian is given by the high-frequency Magnus expansion
	\begin{equation}\label{key}
		\begin{aligned}
			& \hat{H}_{\text{average}} = \hat{H}_{1} +\hat{H}_{2} +\hat{H}_3 +\dots\\
			& \hat{H}_1 = \frac{1}{T} \int^{T}_0 \dif t H(t)\\
			& \hat{H}_2 = \frac{1}{2T} \int^{T}_0 \dif t_2 \int^{t_2}_0 \dif t_1 [\hat{H}(t_2),\hat{H}(t_1)]\,.
		\end{aligned}
	\end{equation}
	
    We can apply Magnus expansion to obtain the average Hamiltonian. In a ``toggling'' frame~\cite{PhysRevX.10.021046} which rotates by $e^{-i\pi/2 \sum^{L}_{i}} \sigma^x_i$ two times in a Floquet period, we neglect higher-order terms and obtain the average Hamiltonian at the lowest order with $H_{\text{average}} \approx \frac{t_3}{T} H_{\text{Int}}+\frac{\epsilon \pi}{T} \sum^L_{i} \sigma^x_i$, which is a $XX$ Hamiltonian with a perturbation transverse field, whose behavior has been calculated in Ref.~\citenum{vogl2019flow}. For $\epsilon =0$ case, it's trivial that $U(T)\sigma^z_i U^\dagger(T) = e^{-i H_\text{Int} t_3} \sigma^z_i e^{i H_\text{Int} t_3} = \sigma^z_i  $, so the quantity $\langle \sum^L_i\sigma^z_i\rangle$ is a conversed quantity that $\langle M(nT)\rangle \sim \langle M(0)\rangle$ under two-period evolution. With the perturbation $\epsilon\sum^L_i \sigma^x_i$, it's a quasi-conversed quantity. In a single period, the $\langle M(nT)\rangle$ has been flipped twice (so we need to measure the system via ``Half Floquet Period Measurement"). However, the expansion is generally believed to be divergent in the many-body system (cannot be expanded into the set of local and few-body operators), so the dynamics of many-body system will gradually extend to the entire many-body Hilbert space. The Floquet Eigenstate thermalization hypothesis (FETH)\cite{ueda2020quantum} asserts the system will eventually seem like an ``infinite temperature state", where the magnetization will totally vanish.
    
    Fortunately, in the the high-frequency regime, the system is governed by the lowest $n$ terms of the average Hamiltonian $\hat{H}_{\text{average}}$. If we truncate the average Hamiltonian to the $n$-th, it will be a Floquet prethermal Hamiltonian and govern the system for a long time. For a time-crystal-like system, it will have a stationary state with a lifetime $e^{C n^*}$ \cite{PhysRevX.7.011026}, where $C$ is a constant and $n^*$ is dependent on the strength of interaction and the the Floquet period $T$.

	\section{\label{sec:DDIS}5. Dynamics of Different Initial States}
\begin{figure*}[htbp]
		\centering
		\includegraphics[width=6.5 in]{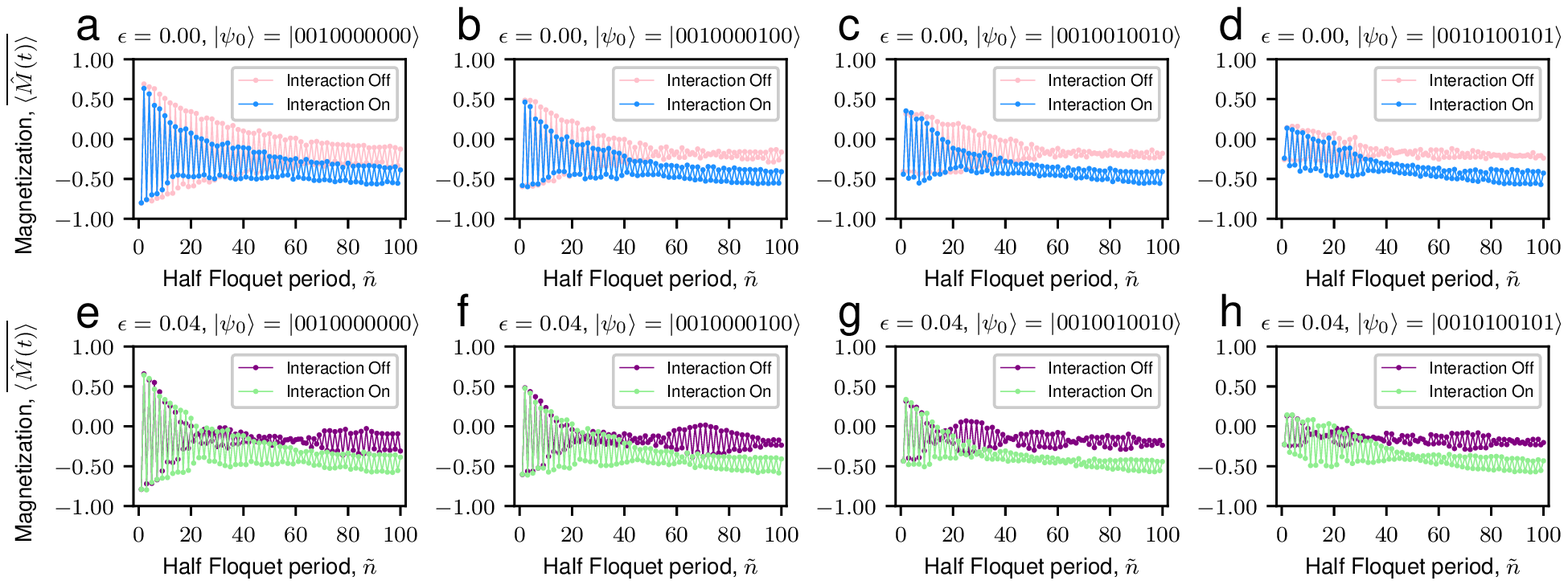}
		\caption{\label{Fig.S1} \textbf{Effect of Different Initial States}. We experimentally probe the magnetization for different initial states $|\psi_0\rangle = |0010000000\rangle $, $ |0010000100\rangle $, $ |0010010010\rangle $ and $ |0010100101\rangle $. \textbf{a}-\textbf{d}, $ \epsilon=0.00 $. Red lines: ``Interaction Off" case. Blue Lines: ``Interaction On" case. \textbf{e}-\textbf{h}, $ \epsilon=0.04 $. Purple lines: ``Interaction Off" case. Green lines: ``Interaction On" case. The dynamics of systems are distinct for ``Interaction Off" and ``Interaction On" cases when a perturbation transverse field exists.}
	\end{figure*}

		To analyze the effect of the initial state on the Floquet matter, we choose different product states to repeat the Floquet protocol in our experiments, shown as Fig. \ref{Fig.S1}. The behavior of the system strongly tends to a prethermal Floquet matter, which can be characterized by the relationship between the lifetime and the temperature of the initial state\cite{PhysRevX.10.021046,PhysRevX.10.011043,kyprianidis2021observation}. Low-temperature initial states still are long-lived while the high-temperature states (multi-excited states) decay quickly. Furthermore, if the external perturbation $ \epsilon $ is applied on the chain, the system will appear the nodes, which is different from the fully polarized state. For the singlet state ($ \langle \hat{M}(0)\rangle =0$), the symmetry and prethermal mechanism both prohibit the DTC-like behavior in the Floquet matter. 

	\section{\label{sec:ED}6. Effects of Decoherence and Leakage}
\begin{figure*}[htbp]
		\centering
		\includegraphics[width=6.5 in]{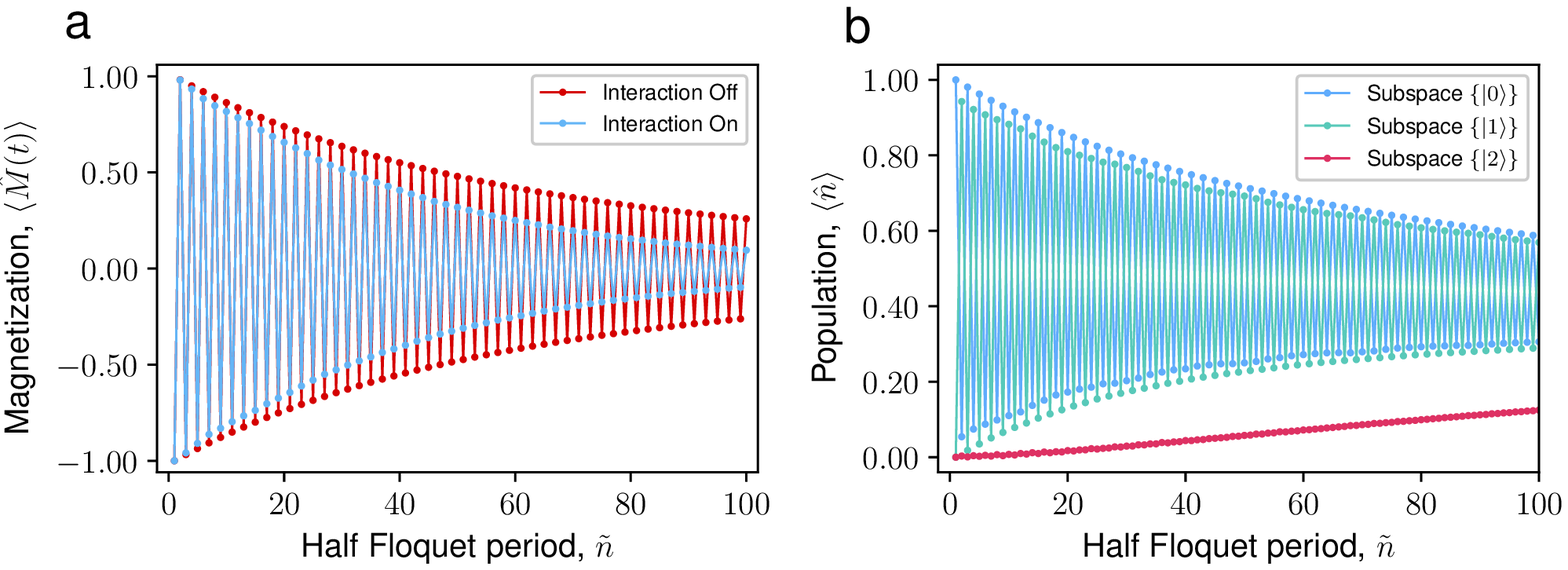}
		\caption{\label{Fig.S2}\textbf{Numerical Simulations for the Effect of Decoherence and Leakage}. \textbf{a,} The time evolution of the magnetization $ \langle \hat{M}(t) \rangle $. The red dots are for ``Interaction Off" case with $ 50 $ complete periods ($ \sim 4000 $~ns) and the blue dots represent  are for ``Interaction On" case ($ \sim 4500 $~ns). \textbf{b,} The occupation will slowly leakage from $|0\rangle$ and $|1\rangle$. After $50$ periods, the population of $|2\rangle$ is about $0.12$. Both the simulation is based on master equation with the effects of decoherence.}
	\end{figure*}

	A major obstacle of the current quantum simulation device is that qubits are vulnerable to the noise induced by inevitable interaction between qubits and the environment, resulting in limited decoherence times. In our scheme, it's essential to distinguish the effect of decoherece and the intrinsic dissipation of the system. In order to estimate the effect of decoherence, we can numerically solve the Lindblad master equation for the density matrix $ \rho $ of qubits,
	\begin{equation}\label{Eq.S3}
		\frac{\dif \rho}{\dif t} = -i[H(t),\rho]+\frac{1}{2}\sum^L_{i=1}\bigg(2\hat{C}_i \rho(t) \hat{C}^\dagger_i -\{\hat{C}^\dagger_i \hat{C}_i,\rho(t) \}\bigg)\,,
	\end{equation}
    where $ \hat{C}_i =\sqrt{\gamma_i} \hat{A}_i $ is the collapse operator. There are two effects of the decoherence (i.e., the energy relaxation effect and the dephasing effect) characterized by the $ T_1 $ and $ T_2^* $ in the Table.~\ref{TableS1}, respectively. as the collapse operators. With Eq.~\ref{Eq.S3}, we can estimate the effect of decoherence in the ``Interaction Off" case and ``Interaction On" case (See the Fig.~\ref{Fig.S2}a). The decay of the magnetization $ \langle \hat{M}(t)\rangle  $ is consistent with the experimental results, shown as Fig.~2.\textbf{c}.
    
    Superconducting transmons are not true two-level systems, but anharmonic oscillators, of which the lowest two levels can be used as a qubit. Leakage errors occur when the transmon system leaves from computational subspace spanned by $|0\rangle$ and  $|1\rangle$.  We utilize DRAG correction to minimize the leakage error of the single qubit gate in the flip stage. In the interaction stage, there are potential leakages mainly due to the energy swap between $|11\rangle$ and $|02\rangle$. Here we simulate the prethermal matter with a $5$-qubit three-level system. When we perform typical dichotomic measurement, the state $|2\rangle$ is apt to be measured as state $|1\rangle$ instead of $|0\rangle$. Therefore the leakage leads to the imbalance of $|0\rangle$ and $|1\rangle$.
    
    \section{\label{sec:NRBO}7. Numerical Results for the Behavior of Observables}
       Here we present the numerics of the spatially-averaged spin magnetization $M(t)$ with the initial state $|0000000000\rangle$ and its spectrum, and the dynamics of the spin glass order parameter (see the Fig.~2\textbf{c}, Fig.3, and Fig.4 in the main text). In the numerical simulations, each point is the average of $20$ sets of disorder.

    \begin{figure*}[htbp]
		\centering
		\includegraphics[width=6.5 in]{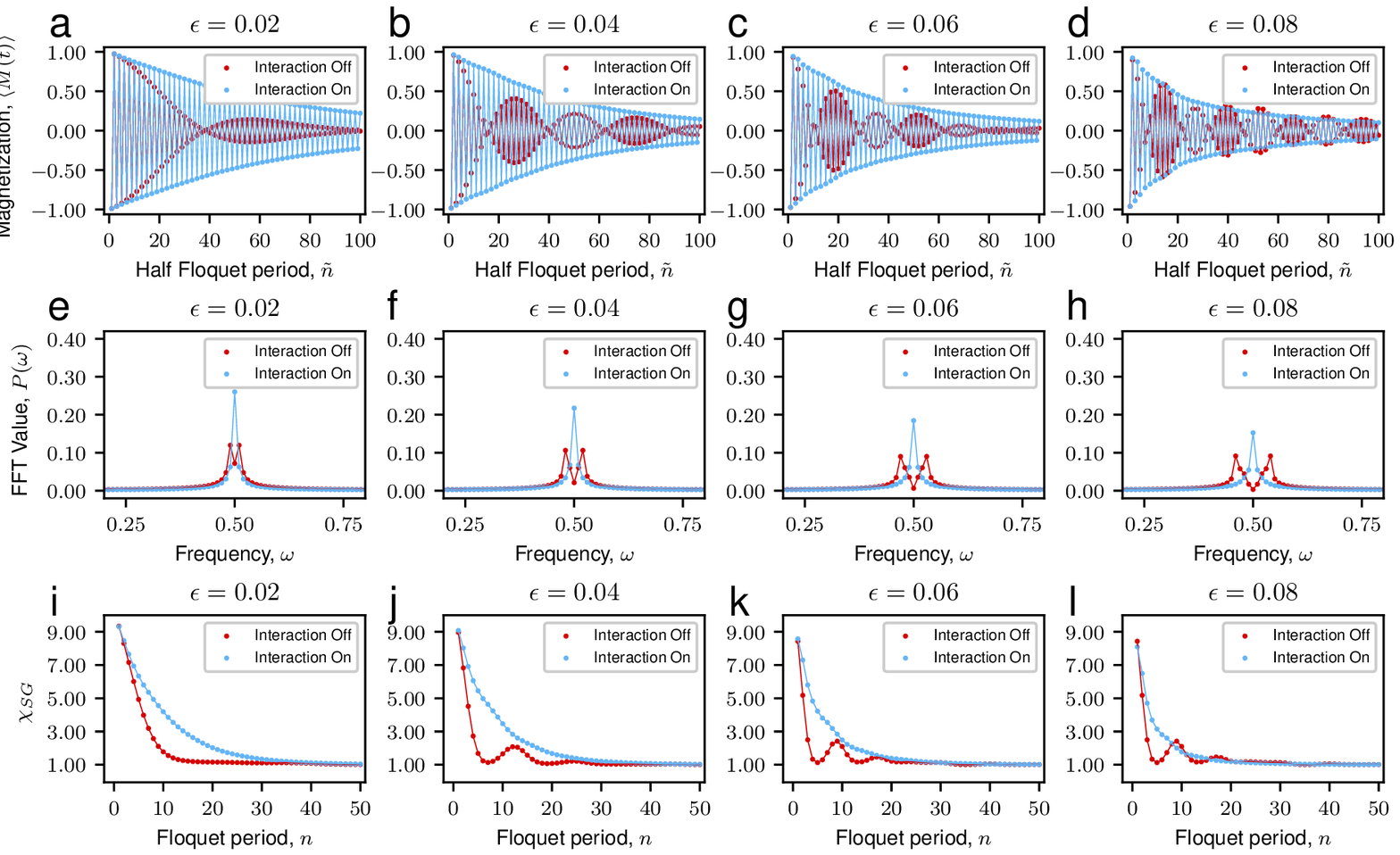}
		\caption{\label{Fig.S3}\textbf{Numerical Simulations for the Behavior of Observables}. \textbf{a}-\textbf{d}, The time evolution of the magnetization $ \langle \hat{M}(t) \rangle $ with various perturbation $\epsilon$. \textbf{e}-\textbf{h}, The Fourier spectra of the magnetization. \textbf{i}-\textbf{l}, The dynamics of the spin glass order parameter.}
	\end{figure*}
	
	\section{\label{sec:NSLTB}8. Disordered Behaviors Of the Non-interacting system}
\begin{figure*}[htbp]
		\centering
		\includegraphics[width=6.5 in]{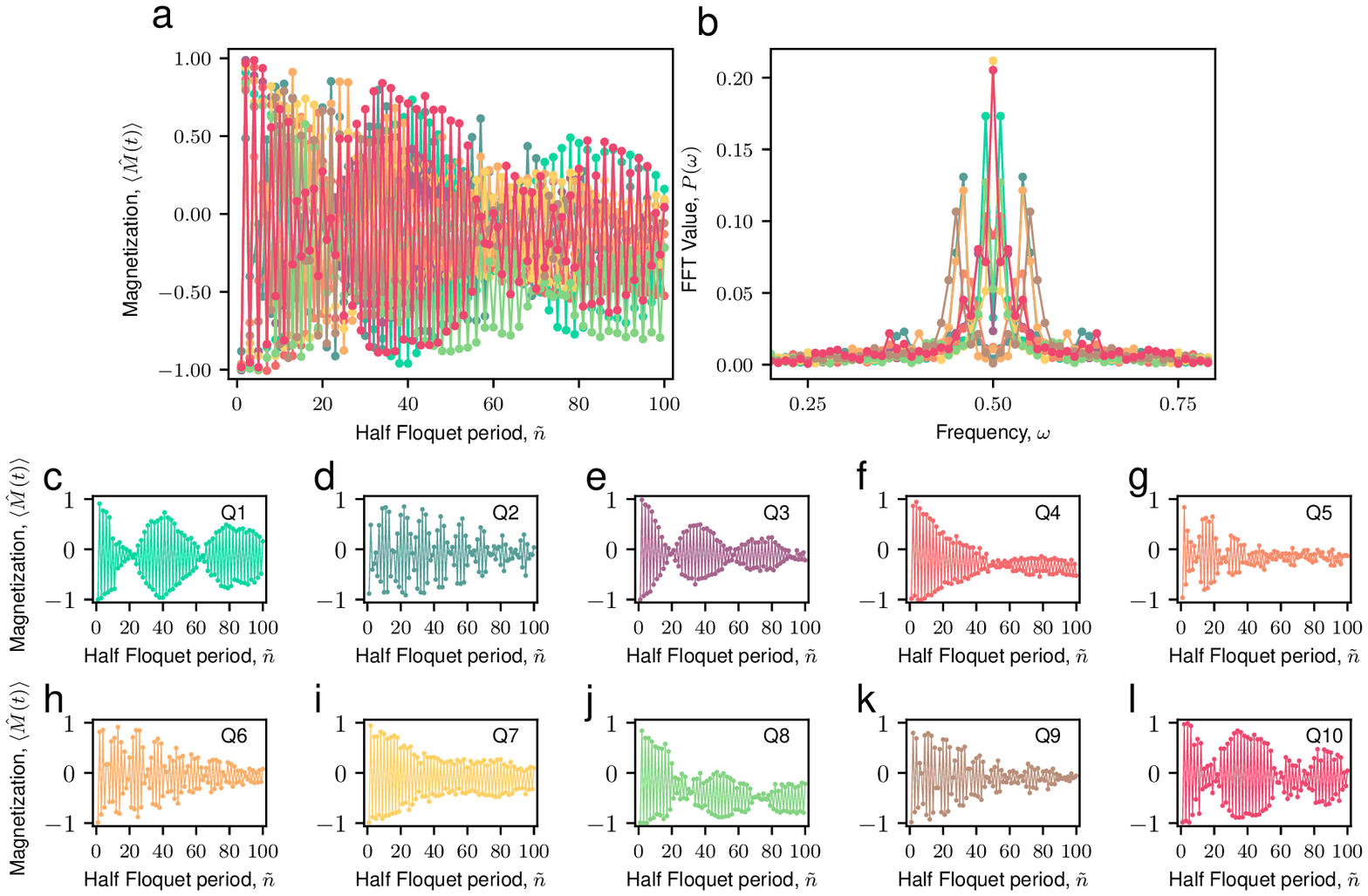}
		\caption{\label{Fig.S4}\textbf{Disordered Behaviors Of the Non-interacting system}. \textbf{a,} Magnetization $\langle \hat{M}\rangle $ of all sites in the Floquet matter with  the external perturbation $\epsilon = 0.10 $. \textbf{b,}. The Fourier spectrum of Magnetization $\langle \hat{M}\rangle $ of all sites in \textbf{a}. \textbf{c-l,} Magnetization $\langle \hat{M}\rangle $ of each site $i$ ($i\in [1,10]$). Each qubit have different dynamics in the presence of the disorder field.}
	\end{figure*}

	 Different from the clean Fourier spectrum in Refs.~\cite{Choi2017,PhysRevLett.120.180603}, there are many peaks in the Fourier spectrum due to the presence of the disorder field. We showed the behaviors of the non-interacting system with the external perturbation $\epsilon = 0.10 $, shown in Fig.~\ref{Fig.S4}. The dynamics of each qubit will be entirely different in the presence of disorder. For some qubits, such as $Q_2$ and $Q_9$, they will oscillate with many beats and the FFT value of $\langle \sigma^z_i\rangle$ will split into two peaks. However, the dynamics of qubit as $Q_7$ will not appear a beat and then the peak in FFT spectrum keeps a single peak. The phenomenon that spins precess with different Larmor rates is due to the random disorders applied on each site, which was reported in Ref.~\citenum{Zhang2017}. For $\epsilon \neq 0$ cases, the central peak in Fig.~3\textbf{a} appears because single peaks exist on some sites (like $Q_4,Q_7$) and some splits are too small (like $Q_1$). Thus, the average FFT value of $\overline{\langle M(t) \rangle} $ will appear many peaks.
    
	\section{\label{sec:NSLTB}9. Numerical Simulations for Long Time Behavior}
\begin{figure*}[htbp]
		\centering
		\includegraphics[width=6.5 in]{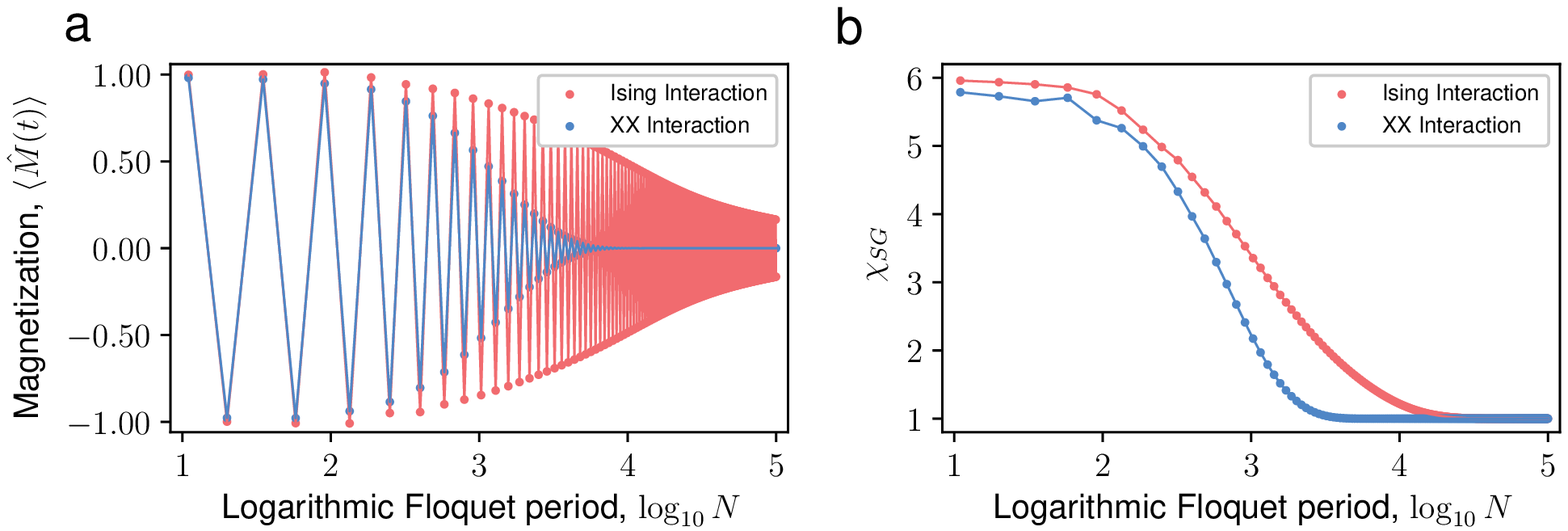}
		\caption{\label{Fig.S5}\textbf{Numerical Simulations for Long Time Behavior}. \textbf{a,} Long time behavior of Floquet matter with $ XX $ interaction and Ising interaction. For the $XX$ interaction (Blue), the system will keep oscillating for $ 10^{3.6} $ Floquet periods. However, the system will keep oscillating for more than $ 10^{5} $ Floquet periods with the Ising Interaction. Prethermal plateau both appear on the dynamics of $M(t)$ in $XX$ interaction and Ising interaction. \textbf{b,} The spin glass order parameter in the prethermal matter. In the $XX$-Interaction case, the order is more fragile and only retains $10^{3.5}$ periods while it has the lifetime about $10^{4.3}$ in the Ising case.}
	\end{figure*}

	In this section, we present more numerical details of the behavior of Floquet matter with $6$ spins under a long time evolution ($ \sim 10^5 $ Floquet periods), shown as Fig.~\ref{Fig.S5}. For the Floquet matter with $ XX $ interaction term, the system will totally decay after $ ~10^{3.6} $ ($ 16000 $) Floquet periods. However, the dynamics of the Floquet matter with Ising interaction will be more robust and last for more than $ 10^{5} $ Floquet periods, whose lifetime is shorter than a discrete time crystal in Floquet system ($\sim 10^{10}$ periods in Ref.~\citenum{PhysRevLett.117.090402}). The difference of robustness between $XX$ interaction and Ising interaction are also shown in the Fig.~\ref{Fig.S5}. The lifetime of the phase with Ising interaction is longer than the $XX$ case.

	\begin{acknowledgments}

	\end{acknowledgments}
	
	\bibliographystyle{apsrev4-2}
	\bibliography{SupM.bib}